\documentclass[preprint,12pt]{elsarticle}

\usepackage{amssymb}

\begin{document}

\begin{frontmatter}



\title{Simulation of Particle Release for Diffusing Alpha-Emitters Radiation Therapy}


\author[1]{Dmytro Fedorchenko\corref{cor1}}
\ead{d.fedoechenko@gmail.com}
\cortext[cor1]{Corresponding author}

\author[1]{Shlomi Alani}
\ead{shlomialani@me.com}

\affiliation[1]{organization={Ziv Medical Center},
            addressline={Derech HaRambam},
            city={Zefat},
            postcode={13100},
            country={Israel}}

\begin{abstract}
We used Monte Carlo simulations to study release of radium-224 daughter nuclei from the seed used for Diffusing Alpha-Emitters Radiation Therapy (DART). Calculated desorption probabilities for polonium-216 (15\%) and lead-212 (12\%) showed that they make a significant contribution to total release from the seed. We also showed that the dose to tissue from decays inside the 10mm long seed exceeds 2.9~Gy for initial radium-224 activity of  111~kBq.

\end{abstract}



\begin{keyword}


Monte Carlo simulation\sep Radium-224\sep Alpha-particle therapy
\end{keyword}

\end{frontmatter}


\section{Introduction}\label{introduction}

Radiation therapy using alpha particles is considered a very promising
approach in treatment of malignant tumors. Alpha particles, having high
linear energy transfer (LET) and a range of 50-100 $\mu$m in human tissue,
provide excellent possibilities for highly localized tumor treatment.
The traditional approach to alpha therapy implies usage of radionuclides
emitting single alpha particle, such as, for example,
\textsuperscript{213}Bi, \textsuperscript{221}At and
\textsuperscript{212}Pb. More advanced techniques called in vivo
generators \cite{borchardtTargetedActinium225Vivo2003} 
use decay chains where several alpha particles are
emitted. Radionuclides \textsuperscript{225}Ac, \textsuperscript{224}Ra
and \textsuperscript{223}Ra with four alpha particles in the decay chain
are considered as prospective candidates for such an approach.

Short alpha particle range requires targeted delivery of the mother
radionuclide to the tumor tissue for the efficient treatment. This could
be achieved by using specific radiolabeled compounds acting as targeting
agents. However, in the case of in vivo generators tumor targeting using
labeled agents is complicated because of nuclear recoil effect
\cite{dekruijffCriticalReviewAlpha2015}. Alpha-decay of the mother nucleus produces a daughter nucleus
with kinetic energy of 100-200 keV which is enough to break chemical
bonds and to leave the targeting agent molecule.

\begin{figure}
	\centering
	\includegraphics{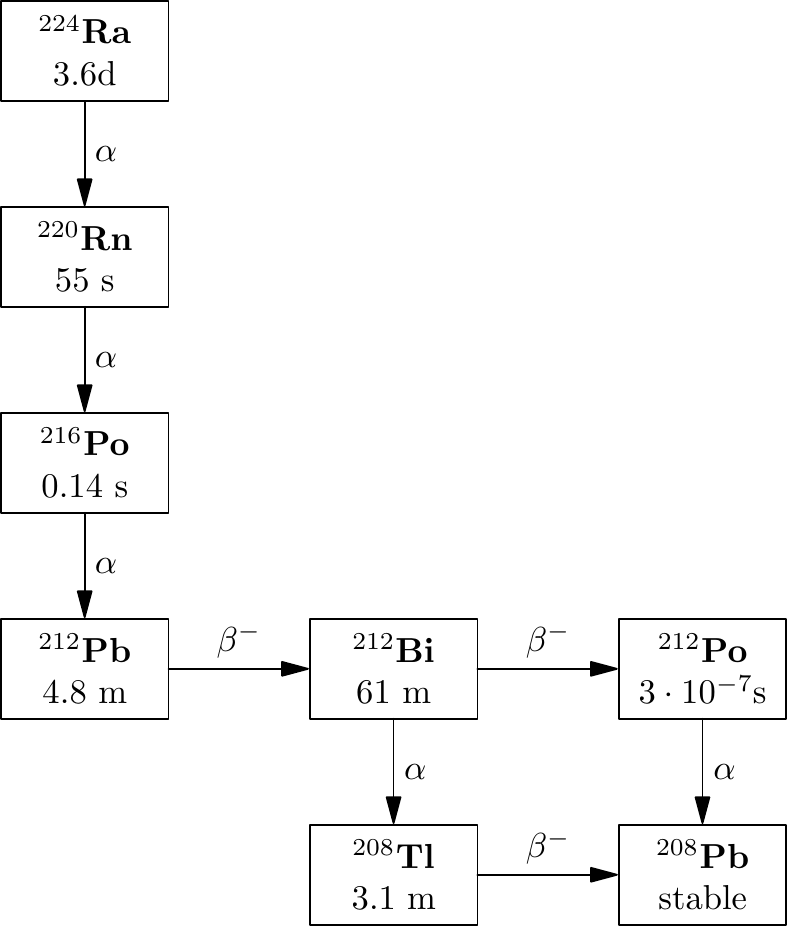}
	\caption{\textsuperscript{224}Ra decay chain\label{fig:decaycahin}}
\end{figure}

The possible solution to the problem is local administration of
alpha-emitting radionuclide directly into the tumor volume. The example
of such an approach is diffusing alpha-emitters radiation therapy (DART)
\cite{Arazi2007,Arazi2010,Arazi2020,cooksLocalControlLung2009,popovtzerInitialSafetyTumor2020}.
According
to this method thin seeds impregnated by \textsuperscript{224}Ra are
placed inside the tumor. Decay of \textsuperscript{224}Ra nucleus
produces alpha particle and \textsuperscript{220}Rn recoil nucleus 
(Figure~\ref{fig:decaycahin}). The recoil nucleus gains enough energy to leave the thin
radium-impregnated surface layer of the DART seed. Diffusion of the
released daughter nuclei forms the treatment volume of several
millimeters with high alpha particles dose. At the same time radiotoxic
radium nuclei remain within the seed. Clinical trials of DART showed
promising results for treatment of several types of tumors
\cite{popovtzerInitialSafetyTumor2020}.

The comprehensive theoretical model describing the long-range dynamics
of recoil nuclei escaping the DART seed was developed by Arazi
\cite{Arazi2020}. The
model is based on the system of diffusion equations with terms
describing nuclei sources and decay. Solution of this system gives
concentrations of the nuclei released from the needle on the distances
of several millimeters.

However, within the macroscopic approach it is impossible to describe
release of the recoil nuclei from the DART seed. This process requires
detailed microscopic description because recoil nuclei emerging in a
thin surface layer undergo only a small number of interatomic
interactions before escaping the seed.

The aim of this work was to study transport of recoil nuclei emerging
from radionuclide decay of \textsuperscript{224}Ra and its daughter
nuclei using Monte Carlo simulation. Our model included simulation of
radioactive decay processes with subsequent detailed simulation of
daughter nuclei and decay products transport.

\section{Monte Carlo simulation
setup}\label{monte-carlo-simulation-setup}

For the simulations of radioactive decay and transport processes we used
GEANT4 simulation toolkit version 11.02 \cite{agostinelliGeant4SimulationToolkit2003,allisonGeant4DevelopmentsApplications2006,allisonRecentDevelopmentsGeant42016}. The toolkit has a very flexible
architecture based on a collection of C++ classes covering various
aspects of particle transport simulation. This allows selection and
customisation of physical models to satisfy requirements of the
particular problem.

Simulation of the recoil nuclei transport in the thin surface layer of
DART seed requires usage of appropriate physical models. The main
physical process responsible for the transport is ion elastic scattering
and only a small number of collisions occur in a thin layer. In this
case one needs an accurate simulation of individual collisions with
realistic interatomic potential. The standard models of ion elastic
scattering used by GEANT4 are based on multiple scattering (MSC)
approximation \cite{ivanchenkoGeant4ModelsSimulation2010}.
This approximation uses the statistical approach valid only when
the moving particle undergoes a significant number of collisions and in
the case of a nanometer-thick layer this condition is not satisfied.

Instead of MSC model we used an accurate single scattering model
developed by Mendenhall and Weller \cite{mendenhallAlgorithmsRapidComputation1991},
and later implemented for GEANT4 toolkit as a G4ScreenedNuclearRecoil class
\cite{mendenhallAlgorithmComputingScreened2005}. This model considers classical pair interactions of the
moving ions with screened Coulomb potential. Simulations of ion
scattering in thin foils and ion implantation showed perfect agreement
of the model with the experimental data \cite{mendenhallAlgorithmComputingScreened2005}.

The DART seed was simulated as a solid cylinder with diameter of 0.3~mm
and length of 10~mm. The seed dimensions correspond to those used for
theoretical and experimental research \cite{Arazi2007}.
Seed material was stainless steel with element composition taken from
the GEANT4 material database: Fe - 74\%, Cr - 18\%, Ni - 8\% .

\section{Results and discussions}\label{results-and-discussions}
\subsection{Simulation of \textsuperscript{220}Rn
release from impregnated layer}\label{simulation-of-220rn-release-from-impregnated-layer}

The key parameter of the DART seed model is thickness of the
radium-impregnated layer. This parameter determines the decay products
release from the DART seed. To determine the realistic thickness value
we used the available data on \textsuperscript{220}Rn release
\cite{Arazi2020}.
According to these experimental studies \textsuperscript{220}Rn
desorption probability is 40$\pm$4\%. This value could be used to estimate
thickness of the impregnated layer by evaluating \textsuperscript{220}Rn
release for different layer thicknesses.

Simulation of the impregnated layer was performed by placing
\textsuperscript{224}Ra ions in random positions within the surface
layer of the DART seed of the given thickness. After that decay of these
was simulated and \textsuperscript{220}Rn were transported in the seed
volume. The desorption probability was calculated as a ratio of
\textsuperscript{220}Rn nuclei crossing the seed surface to the number
of decayed \textsuperscript{224}Ra nuclei.

\begin{figure}
	\centering
	\includegraphics{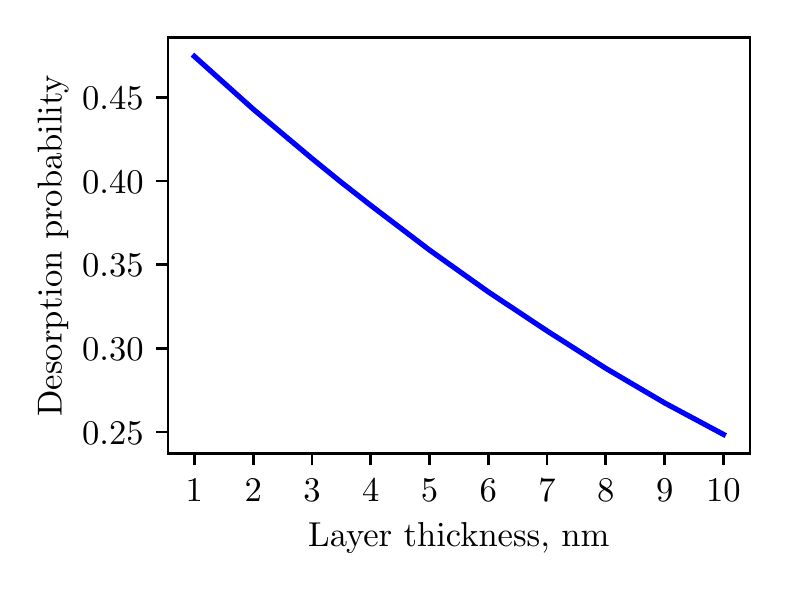}
	\caption{Calculated \textsuperscript{220}Rn desorption probability for
		different thickness of radium-impregnated surface layer of the DART seed\label{fig:rnrelease}}
\end{figure}

The calculated values of \textsuperscript{220}Rn desorption probability
for different thicknesses of the impregnated layer are presented in
Figure~\ref{fig:rnrelease}. According to the obtained dependency \textsuperscript{220}Rn
desorption probability of 40\% corresponds to 3.5~nm thick impregnated
layer. This layer thickness was used for subsequent simulations of the
radioactive decay products transport.

\subsection{Simulation \textsuperscript{224}Ra decay
products transport inside the
seed}{Simulation 224Ra decay products transport inside the seed}\label{simulation-224ra-decay-products-transport-inside-the-seed}
Radon is not the only member of the radium decay chain that is released
from the DART seed. Other nuclei from \textsuperscript{224}Ra decay
chain also gain enough recoil energy to escape the seed body. These
escaped nuclei affect the dose distribution around the seed.

The important factor that determines release of \textsuperscript{224}Ra
decay products from the DART seed is their distribution inside the seed.
This distribution is formed by recoil nuclei traveling from the surface
layer impregnated by \textsuperscript{224}Ra. Part of these nuclei
escapes from the seed, but the remaining part travels away from the
surface. Recoil energy of about 100~keV gained from alpha decays in the
decay chain allows the emerging daughter nuclei to penetrate deeper into
the seed. For beta decay the recoil energy is small and displacements of
the emerging nuclei from the point of decay are small.

\begin{figure}
	\centering
	\includegraphics{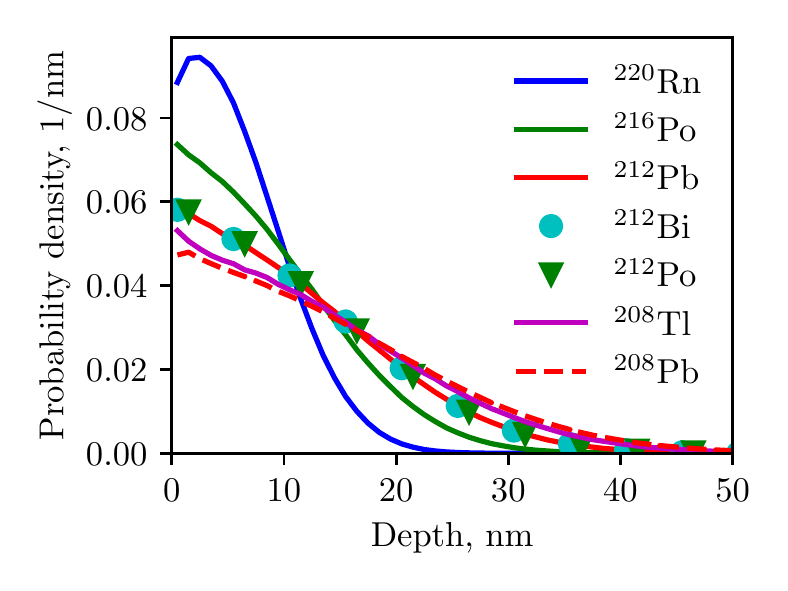}
	\caption{Spatial distribution of radionuclides belonging to
		\textsuperscript{224}Ra decay chain inside the seed.\label{fig:depths}}
\end{figure}

For the simulation we randomly placed \textsuperscript{224}Ra nuclei in
the 3.5~nm thick surface layer, and simulated decays and subsequent
transport of all of its daughter nuclei. Figure~\ref{fig:depths} shows the spatial
distribution of recoil nuclei inside the seed obtained from the Monte
Carlo simulation. These distributions could be characterized by an
average depth at which the recoil nuclei are located in the seed body.
Table~\ref{tab:depths} contains the average depths calculated using the
spatial distributions.

\begin{table}
	\centering
\begin{tabular}{|c|c|}
\hline
Nucleus & Depth, nm\\
\hline
\textsuperscript{220}Rn & 6.278\\
\textsuperscript{216}Po & 8.843\\
\textsuperscript{212}Pb & 11.232\\
\textsuperscript{212}Bi & 11.252\\
\textsuperscript{212}Po & 11.306\\
\textsuperscript{208}Tl & 12.926\\
\textsuperscript{208}Pb & 13.778\\
\hline
\end{tabular}
\caption{Average depths of \textsuperscript{224}Ra daughter nuclei in
		the DART seed\label{tab:depths}}
\end{table}

The calculated distributions show that every next nuclei in
\textsuperscript{224}Ra decay chain penetrate deeper into the seed
forming the elongated distribution tail. While for
\textsuperscript{220}Rn nuclei the distribution spreads up to 25~nm depth
with average depth of 6.3~nm, \textsuperscript{208}Pb nuclei could
travel up to 50~nm from the surface with average depth of 13.8~nm. For
beta decay products \textsuperscript{212}Bi and \textsuperscript{212}Po
displacement from the point of decay is small and their spatial
distributions almost coincide with the distributions of the
corresponding parent nuclei, and have almost equal average depths.

Table~\ref{tab:release} contains desorption probabilities for the
\textsuperscript{224}Ra daughter nuclei obtained from the Monte Carlo
simulation. The desorption probabilities are calculated as ratios of the
number of nuclei of particular sort leaving the seed to the number of
decayed \textsuperscript{224}Ra nuclei. One can see
\textsuperscript{220}Rn has the highest desorption probability of 40\%,
however, desorption probabilities for \textsuperscript{216}Po (15\%) and
\textsuperscript{212}Pb (12\%) are also significant for formation of
dose distribution around the seed.

\begin{table}
\centering	
\begin{tabular}{|c|c|}
\hline
Nucleus & Release fraction\\
\hline
\textsuperscript{220}Rn & 0.398710\\
\textsuperscript{216}Po & 0.149228\\
\textsuperscript{212}Pb & 0.120938\\
\textsuperscript{212}Bi & 0.001048\\
\textsuperscript{212}Po & 0.0022797\\
\textsuperscript{208}Tl & 0.032602\\
\textsuperscript{208}Pb & 0.0765695\\
\hline
\end{tabular}

\caption{Desorption probabilities of \textsuperscript{224}Ra daughter nuclei \label{tab:release} }
\end{table}

The high desorption probability of \textsuperscript{220}Ra nuclei is
determined by the distribution of the parent \textsuperscript{224}Ra
nuclei, which is assumed to be uniform and confined to a 3.5~nm thick
surface layer. For such distribution the average depth is 1.75~nm and
escape probability of emerging \textsuperscript{220}Rn nuclei is rather
high. Distribution of its immediate daughter \textsuperscript{220}Rn has
a much larger depth of 6.3~nm (see Table~\ref{tab:depths}) that results in a
substantially lower desorption probability for \textsuperscript{216}Po.
The average depth of \textsuperscript{216}Po nuclei distribution inside
the seed does not differ considerably from that of
\textsuperscript{220}Rn and both radionuclides have similar desorption
probabilities.

The low recoil energy of beta-decay products \textsuperscript{212}Bi and
\textsuperscript{212}Po results in their short ranges in seed material.
In this case the range defines the thin surface layer from which escape
of the recoil nuclei is possible. From Figure~\ref{fig:depths} one could see that only
a small part of \textsuperscript{212}Bi and \textsuperscript{212}Po are
located in such a surface layer and could actually leave the seed
volume. The corresponding desorption probabilities are very low compared
to those of the alpha decay products. The difference in desorption
probabilities between \textsuperscript{212}Bi and
\textsuperscript{212}Po could be understood from the Q-values of the
corresponding beta decays: 569.9~keV for \textsuperscript{212}Pb decay
and 2252.1~keV for \textsuperscript{212}Bi decay
\cite{brownENDFBVIII8th2018}.
The higher Q-value and higher recoil energy of \textsuperscript{212}Po
nuclei results in higher desorption probability.

Desorption probability for \textsuperscript{208}Tl is low in comparison
to other alpha decay products belonging to the \textsuperscript{224}Ra
decay chain. The reason for this is that the probability of alpha decay
of its parent nucleus is only 35.9\% (Figure~\ref{fig:decaycahin}) and a smaller number of
\textsuperscript{208}Tl nuclei is produced.

Stable radionuclide \textsuperscript{208}Pb is produced by beta decay of
\textsuperscript{208}Tl and alpha decay of \textsuperscript{212}Po. The
recoil energy gained from the latter process determines the desorption
probability for \textsuperscript{208}Pb, because the probability of
\textsuperscript{208}Pb nuclei escaping after beta decay is very small.
The parent radionuclide \textsuperscript{212}Po is produced with the
probability of 64.1\%, and this is also the fraction of
\textsuperscript{208}Pb nuclei produced from alpha decay and having a
reasonable escape probability. This results in the relatively low
\textsuperscript{208}Pb desorption probability compared to desorption
probabilities of other decay chain members produced from alpha decay.

\subsection{Simulation of recoil nuclei transport in
tissue}\label{simulation-of-recoil-nuclei-transport-in-tissue}

\begin{figure}
	\centering
	\includegraphics{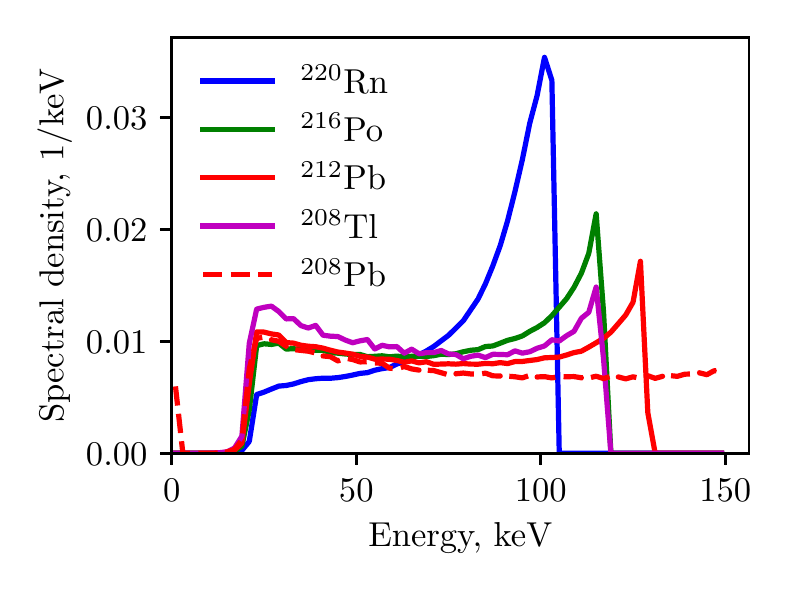}
	\caption{Spectra of the recoil nuclei escaping the DART seed\label{fig:spectra}}
\end{figure}

The range of recoil nuclei in tissue is determined by their spectra on
the seed-tissue border. Figure~\ref{fig:spectra} shows calculated spectra of the recoil
nuclei that escape the seed and enter the surrounding tissue. The figure
does not include beta decay products \textsuperscript{212}Bi and
\textsuperscript{212}Po because due to low recoil energy they have a
very short range in tissue.

The calculated spectrum for \textsuperscript{220}Rn exhibits a
pronounced peak close to the recoil energy of 103.4~keV that the radon
nucleus gains from alpha decay of \textsuperscript{224}Ra. Such a
distribution shows that an essential part of \textsuperscript{220}Rn
nuclei escapes from the seed without scattering and accompanying energy
loss. The physical reason for this is that parent
\textsuperscript{224}Ra nuclei are concentrated in the thin surface
layer and emerging radon nuclei leave the seed without interacting with
nuclei of seed material.

For other alpha decay products the peak around the recoil energy is less
pronounced. This is the result of elastic scattering on the nuclei of
the seed material that effectively lowers the energy of the moving
nucleus. For the \textsuperscript{208}Pb spectrum there is also a small
peak at lower energies. This peak originates from
\textsuperscript{208}Pb nuclei emerging from beta decay of
\textsuperscript{208}Tl and having low recoil energy.

Recoil nuclei escaping the seed travel for some distance in tissue
before they slow down to thermal energies. The further transport of
these nuclei and their decay products in tissue is due to diffusive
processes \cite{Arazi2020}. We did not consider diffusion and our simulation of nuclei
transport in tissue included only the recoil nuclei emerging in the seed
volume and entering the tissue. Also, during simulation radioactive
decay of recoil nuclei traveling in tissue was prohibited. This is
physically reasonable because half-life periods of radionuclides
belonging to \textsuperscript{224}Ra decay chain are considerably longer
than typical times of recoil nuclei slowing down that could be estimated
around 10\textsuperscript{-15} s.

\begin{figure}
	\centering
	\includegraphics{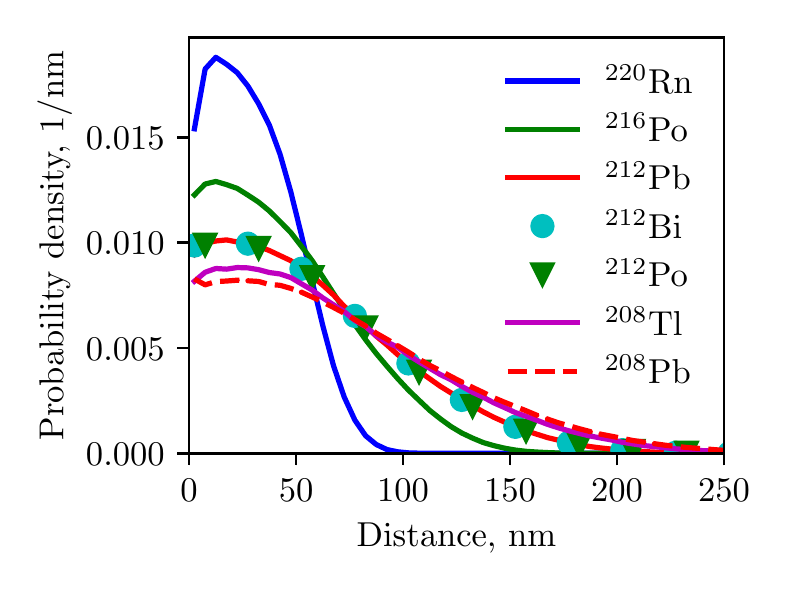}
	\caption{Distribution of recoil nuclei ranges in tissue\label{fig:ranges}}
\end{figure}

Figure~\ref{fig:ranges} shows distributions of recoil nuclei ranges in tissue obtained
from Monte Carlo simulation. The corresponding average range values are
presented in Table~\ref{tab:ranges}. One can see that recoil energy allows nuclei
escaping the seed to travel to about 100 nm in tissue. The average
ranges for all radionuclides from the decay chain except for
\textsuperscript{208}Pb have very close values of about approximately 30~nm. 
The obtained value is much smaller than the seed radius of 0.15~mm.
This shows that for macroscopic models describing diffusion of
\textsuperscript{224}Ra decay products one can neglect the recoil nuclei
penetration into the tissue and assume that they decay directly on the
seed surface. However, if we are interested in the dose distribution
near the seed surface such an assumption will be illegal.

\begin{table}
	\centering
	\begin{tabular}{|c|c|}
\hline
Nucleus & Distance, nm\tabularnewline
\hline
\textsuperscript{220}Rn & 30.3\\
\textsuperscript{216}Po & 30.7\\
\textsuperscript{212}Pb & 32.2\\
\textsuperscript{208}Tl & 29.2\\
\textsuperscript{208}Pb & 39.2\\
\hline
\end{tabular}

\caption{Average recoil nuclei ranges in tissue\label{tab:ranges}}
\end{table}

Dose distribution around the seed is formed by coupled processes of
radioactive decay and diffusion. The unstable recoil nuclei escaped from
the seed are transported by diffusion and undergo radioactive decay
during movement. The characteristic values of diffusion lengths obtained
from macroscopic models are 0.3~mm for \textsuperscript{220}Rn and 0.6~mm 
for \textsuperscript{212}Pb, and radial dimensions of the region with
therapeutic levels of alpha-particle dose is about 5-10 diffusion
lengths \cite{Arazi2020}.

At the same time \textsuperscript{224}Ra and essential part of its decay
products remain inside the seed. Alpha particles emerging from decays of
these nuclei form dose distribution in the region close to the seed
surface. This distribution could not be calculated within the
macroscopic approach, so we used Monte Carlo simulation to calculate
alpha-particles dose from decays inside the seed.

\begin{figure}
	\centering
	\includegraphics{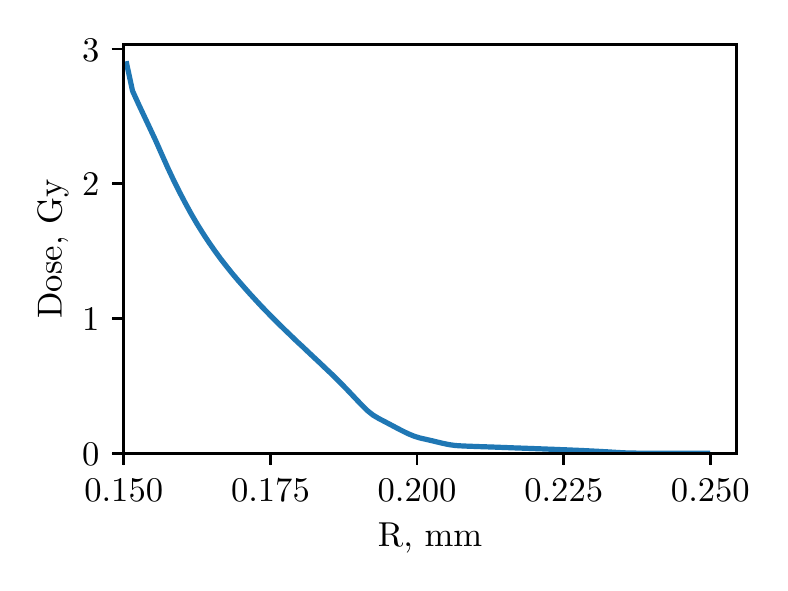}
	\caption{Dose distribution around the seed with radius of 0.15~mm from
		decays inside the seed, where R is radial distance from the seed axis. The
		initial \textsuperscript{224}Ra activity is 3~$\mu$Ci (111 kBq) and
		irradiation time is 10 days\label{fig:doze}}
\end{figure}

Radial dose distribution from alpha particles emerging inside the seed
is shown in Figure~\ref{fig:doze}. This distribution was calculated for a 1 cm long
seed with initial activity of 3~$\mu$Ci (111~kBq) and irradiation time of 10
days. The region where doze is formed by alpha particles leaving the
seed extends approximately to 0.2~mm from the needle axis (0.05~mm from
the surface). The size of this region is considerably smaller than the
size of the treatment region which is around several millimeters
\cite{Arazi2007}.

The calculated dose distribution shows that close to the seed
alpha-particles dose reaches the value of 2.9~Gy. Average dose in the
region irradiated by alpha particles from the seed exceeds 0.49~Gy. This
should be compared to the reference dose of 10 Gy used to define size of
the treatment region \cite{Arazi2007}.
One can see that alpha decays inside the DART seed make an essential
contribution to the total dose and should be accounted for during
treatment planning.

\section{Conclusions}\label{conclusions}

In this paper we examined the release of \textsuperscript{224}Ra decay
products from the DART seed due to nuclear recoil effect. Using Monte
Carlo simulation we calculated their desorption probabilities and showed
that beside \textsuperscript{220}Rn with desorption probability of 40\%,
release of \textsuperscript{216}Po and \textsuperscript{212}Pb with
desorption probabilities of 15\% and 12\% respectively makes noticeable
contribution to the doze to the tissue surrounding the seed. From the
simulation we also obtained distribution of \textsuperscript{224}Ra
decay products remaining inside the seed and calculated dose to tissue
from decays of these nuclei. This dose could exceed 2.9~Gy close to the
seed surface for initial seed activity of 3~$\mu$Ci (111~kBq) and
irradiation time of 10 days. Under these conditions the average dose to
the tissue layer with thickness of 0.05~mm exceeds 0.49~Gy.

The obtained results complement the existing macroscopic theory for the
recoil nuclei transport \cite{Arazi2020}. They
could be used for development of the improved macroscopic theory
describing dynamics of \textsuperscript{224}Ra decay products in tissue.

\bigskip
This research did not receive any specific grant from funding agencies in the public, commercial, or not-for-profit sectors.

\bibliographystyle{elsarticle-num}
\bibliography{article.bib}





\end{document}